# Green-Roof Energy Performance in New Zealand's Present and Future Climate Condition (2050)

**Abdollah Baghaei Daemei**

Building Performance Analysis Lab, Tech Innovation Experts, Auckland, New Zealand

Abdollah.bagheai@tinx.co.nz

**Abstract.** Green roofs are increasingly promoted as nature-based measures for reducing building energy demand, yet their performance in New Zealand's oceanic climates and under future weather remains insufficiently quantified. This condensed study compares an extensive green roof with a conventional bare roof on a standardized single-storey dwelling in Auckland, Christchurch, and Wellington. Dynamic annual simulations were conducted in DesignBuilder/EnergyPlus using present-day EnergyPlus Weather files and 2050 weather files generated with CCWorldWeatherGen. All non-roof building parameters were held constant so that differences in total fuel consumption (TFC) for heating and cooling could be attributed to the roof system. Under present weather, annual TFC decreased by approximately 3.1% in Auckland, 2.3% in Christchurch, and 1.5% in Wellington. Under 2050 weather, the corresponding reductions were 3.3%, 2.6%, and 1.1%. Summer benefits were larger in Auckland and Christchurch, reaching about 9.0% and 5.6%, respectively, in 2050, but remained marginal in Wellington. The findings show that green roofs can provide modest annual energy savings in oceanic climates, with stronger value as a summer heat-mitigation measure in warmer locations. Performance is strongly climate-dependent and should not be generalized without local simulation or field validation.

## 1. Introduction

Buildings must respond simultaneously to rising energy demand, urban warming, and climate-change-driven shifts in heating and cooling loads [1, 2]. Nature-based building-envelope strategies are attractive because they can moderate heat transfer while also contributing to stormwater management, biodiversity, and urban environmental quality [3]. Among these strategies, green roofs introduce vegetation and a growing substrate above the roof assembly. Their thermal influence arises from added resistance and heat capacity, shading, evapotranspiration, and altered surface radiation exchange [4]. The magnitude and direction of the resulting energy benefit, however, depend on climate, roof construction, vegetation properties, moisture availability, building form, internal loads, and HVAC operation [5].

Evidence from tropical, arid, temperate, continental, and cold climates confirms that green roofs can reduce roof-surface temperature and cooling demand, but annual energy savings are often less consistent than peak-period benefits [2-5]. In heating-dominated climates, the additional roof layer may reduce winter heat losses, while in mild or windy locations the net annual effect can be small. Climate change further complicates design decisions because warmer future weather can decrease heating demand, increase cooling demand, and change the seasonal value of passive technologies. A measure that appears marginal under current weather may become more useful during future summer conditions, whereas a solution optimized for one city may perform poorly in another [6].

New Zealand provides a relevant test context. Auckland, Christchurch, and Wellington are all commonly classified within the broad oceanic climate family, yet they differ substantially in temperature, humidity, solar exposure, and wind. Auckland is warmer and more humid; Christchurch has larger seasonal temperature variation and colder winters; Wellington is temperate but persistently windy. Previous New Zealand research has examined climate-change effects on residential heating and cooling demand and has identified the need for climate-responsive energy standards [7]. However, direct comparisons of green-roof energy performance across these cities, including a future-weather scenario, have been limited.

This study therefore addresses two questions: (1) how much can an extensive green roof reduce annual and summertime heating and cooling energy use relative to a conventional bare roof in Auckland, Christchurch, and Wellington; and (2) how does this performance change under projected 2050 weather? The contribution is a controlled cross-city comparison using one building geometry and one roof specification, allowing climatic effects to be examined without changes in building form or operation.

## 2. Methods

A standardized single-storey New Zealand dwelling was modelled in DesignBuilder, which uses the EnergyPlus simulation engine. The house has approximately 182.6 m2 of floor area and 203.8 m2 of roof area. It was selected without an attic or ventilated roof cavity so that the roof assembly had a direct thermal relationship with the occupied zones. The same geometry, orientation, glazing, occupancy assumptions, internal loads, and split-system HVAC configuration were applied in all simulations.

## 2. Methods (continued)

Two roof alternatives were evaluated. The reference case used a conventional pitched metal roof assembly. The intervention case added an extensive vegetated system comprising vegetation, growing medium, waterproofing and supporting layers. The modelled green-roof parameters included a plant height of 0.10 m, leaf area index of 2.7, leaf reflectivity of 0.22, leaf emissivity of 0.95, minimum stomatal resistance of 100 s/m, and initial volumetric moisture content of 0.15. The effective roof U-value used in the model was approximately 0.486 W/m2K. These values represent a generic extensive roof rather than a site-specific plant community.

Present-day simulations used location-specific EPW files for Auckland, Christchurch, and Wellington. Future weather files for 2050 were generated with CCWorldWeatherGen, which morphs baseline weather using the HadCM3 climate model under the IPCC A2 emissions pathway [7,8]. Each model was simulated from 1 January to 31 December. The primary outcome was TFC for space heating and cooling, expressed in Wh/m2. Annual results were evaluated together with summer-period results because green-roof cooling effects are expected to be most visible during warm conditions. The analysis is comparative: savings were calculated as the percentage reduction in TFC for the green roof relative to the bare roof [8].

## 3. Results

Across all three cities and both weather periods, the green roof produced lower annual TFC than the bare roof, but the magnitude varied materially by climate. Under present weather, Auckland's annual TFC decreased from 58,599 to 56,757 Wh/m2, a reduction of 3.1%. Christchurch decreased from 72,787 to 71,125 Wh/m2, equivalent to 2.3%. Wellington decreased from 62,794 to 61,825 Wh/m2, or 1.5%. These values show that the intervention was beneficial in all locations but did not generate large whole-year reductions.

**Table 1.** Summary of the results

| City | Annual saving, present | Annual saving, 2050 | Summer saving, present | Summer saving, 2050 |
|---|---|---|---|---|
| Auckland | 3.1% | 3.3% | 8.5% | 9.0% |
| Christchurch | 2.3% | 2.6% | 4.5% | 5.6% |
| Wellington | 1.5% | 1.1% | 1.5% | 0.2% |

*Table 1. Relative reduction in heating and cooling TFC with the green roof.*

Under 2050 weather, Auckland's bare-roof TFC was 59,074 Wh/m2 and the green-roof value was 57,121 Wh/m2, corresponding to a 3.3% saving. Christchurch decreased from 67,481 to 65,704 Wh/m2, a 2.6% saving. Wellington decreased from 59,273 to 58,634 Wh/m2, only 1.1%. Thus, projected warming did not fundamentally change the annual ranking: Auckland retained the largest benefit, Christchurch remained intermediate, and Wellington showed the weakest response.

The summer analysis revealed a clearer climate signal. In Auckland, the green roof reduced summer TFC by approximately 8.5% under present weather and 9.0% in 2050. Christchurch increased from about 4.5% to 5.6%. Wellington showed only 1.5% under present weather and approximately 0.2% in 2050. The pronounced summer performance in Auckland and Christchurch suggests that vegetation and substrate most effectively offset roof heat gains when warm-period loads are sufficiently high. Wellington's small response indicates that strong wind, lower solar-driven cooling demand, and the balance between heating and cooling conditions can limit the energy value of the same roof specification.

## 4. Discussion

The results support a cautious interpretation. Green roofs are not a high-impact annual energy retrofit in every oceanic climate. The annual reductions of roughly 1-3% are useful but modest, particularly when considered against installation cost, structural requirements, waterproofing risk, irrigation, and maintenance. Claims that a green roof will substantially reduce household energy use are therefore not defensible without project-specific evidence. Its strongest energy rationale in this study is seasonal: the intervention reduced summer demand more clearly in warmer Auckland and, to a lesser extent, Christchurch.

The 2050 simulations indicate that climate change may strengthen the cooling-related value of green roofs in locations where summer loads increase. Auckland and Christchurch both showed slightly larger annual and summer savings in 2050. Wellington did not. This divergence is important for policy and design: a national prescription based only on the label 'oceanic climate' would be technically weak. Green-roof guidance should instead consider local weather, roof exposure, plant and substrate design, moisture availability, HVAC system, and whether the project objective is annual energy reduction, peak-load management, stormwater control, biodiversity, or urban heat mitigation.

## 4. Discussion (continued)

The controlled modelling approach improves internal comparability, but it also limits external validity. Only one detached dwelling, one pitched-roof geometry, one green-roof specification, and one HVAC arrangement were tested. The model did not compare alternative substrate depths, irrigation schedules, plant species, ageing effects, roof orientations, flat roofs, apartments, schools, or commercial buildings. Future weather was generated from a single modelling pathway rather than an ensemble of contemporary climate projections. In addition, simulation results were not calibrated against measured roof temperatures, heat fluxes, indoor conditions, soil moisture, or whole-building energy data. The reported percentages should therefore be interpreted as scenario-specific estimates rather than universal performance coefficients.

The next research step should be experimental validation in at least two contrasting New Zealand climates, ideally Auckland and Christchurch. Instrumented test roofs should measure surface temperature, substrate moisture, heat flux, indoor operative temperature, and HVAC electricity at sub-hourly resolution. A factorial simulation campaign could then test substrate depth, leaf area index, irrigation, insulation level, roof slope, and future climate ensembles. Economic and embodied-carbon analysis is also required. A roof that saves 2-3% of annual HVAC energy may still be justified when stormwater, biodiversity, roof-life extension, and neighbourhood heat reduction are valued jointly, but the energy benefit alone may not support the investment.

## 5. Conclusion

A generic extensive green roof reduced simulated heating and cooling TFC in Auckland, Christchurch, and Wellington under both present and 2050 weather. Annual savings were modest, ranging from approximately 1.1% to 3.3%. Summer savings were more substantial in Auckland and Christchurch, reaching about 9.0% and 5.6% in 2050, but remained negligible in Wellington. The central conclusion is not that green roofs are universally energy efficient; rather, their performance is climate- and context-dependent. For New Zealand, green roofs should be evaluated as multi-benefit infrastructure and designed using local simulation, field measurements, and explicit project objectives.